# Evolution and Mutations of Beta 2 Microglobulin

J. C. Phillips

Dept. of Physics and Astronomy, Rutgers University, Piscataway, N. J., 08854

## Abstract

Here we examine the evolution of beta-2 microglobulin in terms of its hydropathic shapes, a theoretical construct that has revealed important trends. The dynamics of many proteins are largely driven by interactions between the protein itself and the thin water film that covers it. β2m constitutes the basic building unit of the immunoglobulin superfamily; the evolution of its amino acid sequences from chickens to mice to humans provides new information about its multiple functions. Our hydrodynamic method involves concepts of topological shape evolution towards a critical point for optimized functions. The results are in excellent agreement with experiment for the details of the mouse-human evolution, as well as both the dangerous natural amyloid aggregation mutation D76N, and six other DN test mutations.

## Introduction

Because beta-2 microglobulin is not only the basic globulin, but also small (99 amino acids), it has been the subject of many studies, including its evolution from chicken to human [1]. Studies showed that chicken β2m exhibits a lower melting temperature than human β2m, and the H/D exchange behavior observed by infrared spectroscopy indicates a more flexible structure of the former protein. It was also noted that there are more glycine residues in chicken β2m than in human β2m (5 versus 3). These residues could have an effect on the stability of the former molecule, since glycine residues are able to contribute to main chain flexibility. However, mouse has only 2 glycines; it is difficult, even in beta-2 microglobulin, to prove positive evolution by counting residues, or by making macroscopic measurements. Here we will find monotonic positive hydropathic evolution from chicken (49% Blast identity to human) through mouse (70% Blast identity to human) to human beta-2 microglobulin.

D76N is the only natural variant of human beta-2 microglobulin so far identifiedContrary to the wt protein, this mutant readily forms amyloid fibers in physiological conditions, leading to a systemic and severe amyloidosis.  NMR studies and X-ray crystallography have shown that the structures of the 99-residue protein beta(2)-microglobulin (beta(2)m) and its more aggregation-prone variant, D76N, are indistinguishable.  Although the Asp76Asn mutant has been extensively characterized [2-5], the molecular bases of its instability and aggregation propensity are still being studied [6].  The present phase-transition model uses the same methods used to study the monotonic  evolution of a wide range of proteins, including the 1200 amino acid spikes of CoV-1,2, its minor variants, and the major new one, Omicron [7-11].

Methods

The results presented here involve only the amino acid sequences obtained from Uniprot (no new data).  At first this appears to be impossible, but the phase transition method has been successful for many proteins, starting with hen egg white [12].  The data underlying all applications of the method are structural, derived from studying the hydropathic curvatures of > 5000 protein segments of lengths L with $9 \leq L \leq 35$ [13].  Their analysis led to the discovery of 20 amino acid hydropathicity parameters $\Psi(aa)$.  These parameters describe long-range interactions of amino acids with the water films that have shaped all proteins (hence their broad applicability).  The existence of such parameters for second-order phase transitions has been known in principle for decades [14,15], but in general they are not easily measured because they describe only weak interactions very close to phase transition critical points.  It has turned out that evolution has created such data by natural selection, and thus the structural protein data base (and especially protein sequences) already contains a great deal of information, whose exploration has just begun.

The next step is to identify a sliding window length W over which we average $\Psi(aa)$ to obtain the matrix $\Psi(aa,W)$.  One suspects that the best W will also lie between 9 and 35, and that it will be smaller for smaller proteins.  The average protein size is ~ 300 amino acids.  Beta-2 microglobulin is unusually small, so one can guess that the best value of W will be 9.  Our calculations used W = 7, 9, and 11, and W = 9 gave the best results, which are shown.  By comparison, the coronavirus spike, with 1200 amino acids, gave best results with W = 39 [10].

Results

The hydroprofiles Ψ(aa,9) are shown for h Peak 2 is higher in human, and is almost ,e,uman, mouse and chicken.  There are three numbered hydrophobic peaks near 25, 65 and 85.  Peak 1 is almost unchanged by evolution from chicken to human, while peaks 2 and 3 are enhanced in human.  Note also the large hydrophilic valley between 40 and 60, which is deepened by evolution.  This valley acts as a hydrodynamic hinge, allowing rapid bending of peaks 2 and 3, relative to peak 1.

The details of the approach of beta-2 microglobulin to its nearly critical human shape are clearer in Fig. 2, which includes only mouse and human.  Peak 2 is higher in human, and is almost level with peak 1.  Studies of other proteins (most recently, CoV spikes [7-11]) have shown that edge leveling enables dynamic synchronization of interfaces, a concept that has applications to aligning many other shaped interfaces [16].  Here the peak edges of mouse are 1: 188.4, and 2: 179.4, and in human, 1: 189.8, and 2: 191.7, the difference has decreased by more than a factor of 4.  Although peak 3 is not aligned with peaks 1 and 2, it has strengthened substantially from mouse to human.  Hydroneutral is ~ 155, so relative to hydroneutal, the increase of peak 3 in humans has made it 4 times more hydrophobic than peak 3 in mouse.

Multiple disulfide bonds can interfere with water-film stabilized protein shapes, while a single disulfide bond does not [9].  Chicken, mouse and human beta-2 microglobulins contain a single disulfide bond, between sites 25 and 80, or hydrophobic peaks 1 and 3.  This connection helps peak 3 have a stabilizing effect, assisting peaks 1 and 2.  Meanwhile, the hydrophilic edges 4 – 6 are more nearly aligned in human (average deviation, 2.4) than in mouse (average deviation, 5.7).

D76N is the only natural variant of human beta-2 microglobulin (beta 2m).  It has been extensively characterized, but the molecular and structural determinants of its peculiar properties remained elusive.  Both the high resolution crystal structure and the NMR analysis of the mutant did not reveal any significant structural features that could explain the D76N mutant striking properties.  Contrary to the wt protein, this mutant readily forms amyloid fibres in physiological conditions, leading to a systemic and severe amyloidosis.  Molecular dynamics simulations suggested that destabilization of the outer strands of D76N beta 2m accounts for the increased aggregation propensity [5].  The hinges of the outer strands are the hydrophilic edges 4-6 in Fig. 2.  Mutating a single amino acid has only small effects, so Fig. 3 compares the effects of D76N on human $\Psi$(aa,9). The mutation brings hydrophilic (outer) edge 6 (116.3, wt; 119.2, with D76N) into better alignment with edge 4 (121.3): the mutated difference is < half the wild type.

There is a second test for the importance of edge leveling [3]. Other D-to-N mutants (D34N, D38N, D53N, D59N, D96N and D98N) were characterized in terms of thermodynamic stability and aggregation propensity. None showed the dramatic drop in melting temperature (relative to the wt protein) as observed for D76N. Consistently, none of these variants displayed any increase in aggregation propensity. When we compare these sites to the $\Psi$(aa,9) profiles in Fig. 2, we find that only one falls near an edge: D38N lies near the hydrophilic edge 5, but it actually reduces leveling with edges 4 and 6. Among the other six D-to-N mutants only the structure of the D38N mutant displays evident conformational changes relative to wt $\beta$2m (Fig. 5 of [18]). Overall, the match between edge leveling of MZ $\Psi$(aa,9) and the data of [3] is amazingly close.

There are several subtle points included here. One is the choice of W = 9. This value of W makes the window centered on 76 just reach the edge at 73 (see Fig. 3). Another is the choice of the MZ (second order phase transition) scale, instead of the standard KD (first order phase transition, [8,17]) scale. The differences are interesting, and so the results for $\Psi$(aa,9) with the KD scale are shown in Fig. 4. Thus the differences between peaks 1 and 2 with the KD scale for $\Psi$(aa,9) mouse and human is larger than with the MZ scale, and mouse even is slightly smaller than human, inconsistent with the edge leveling effects of evolution [7 -10,18].

Discussion

The molecular dynamics simulations of [5] are among the most successful MDS known to us. Their success was possible because beta 2m is small (100 aa) compared to the average protein (300 aa). The phase transition method works equally well in describing the evolution of proteins of any size, even 1200 aa in the case of CoV spikes [7 - 11]. The key to the success of the phase transition method is the 20 fractals discovered only recently [13]. Indeed, the proliferation of > 100 hydropathicity scales before 2000 had a discouraging effect, and has led many chemists to conclude that all such scales are only qualitative [19,20].

We are now well into the 21[st] century, and because of [9] the quantitative significance of hydropathicity scales has changed [21]. All the sequence data used here are available online for many proteins through Uniprot, while the MZ and KD scales are in [8]. One can easily construct the Ψ(aa,W) matrices, or more simply, use EXCEL. An EXCEL Macro is available from the author on request.

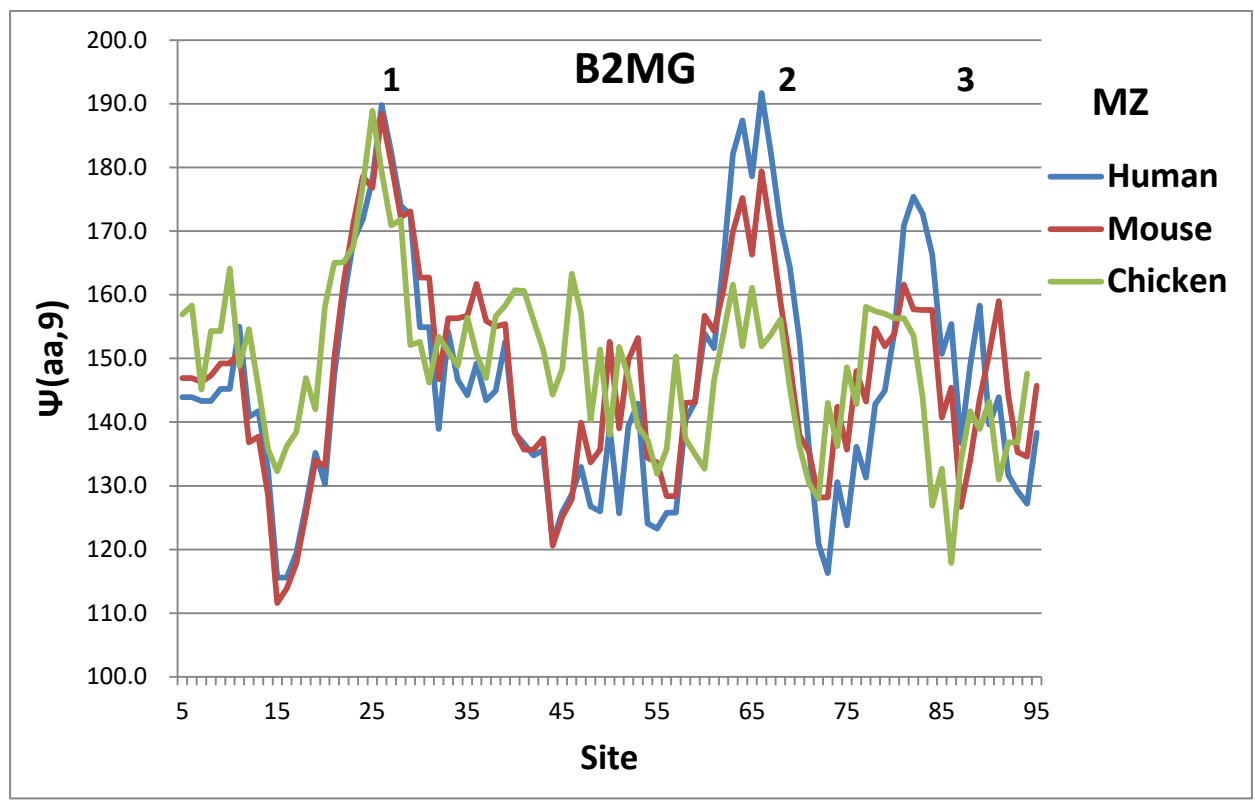

Fig. 1. Hydroprofiles of human, mouse and chicken, using the MZ (second-order phase transition) scale [13] with W = 9.

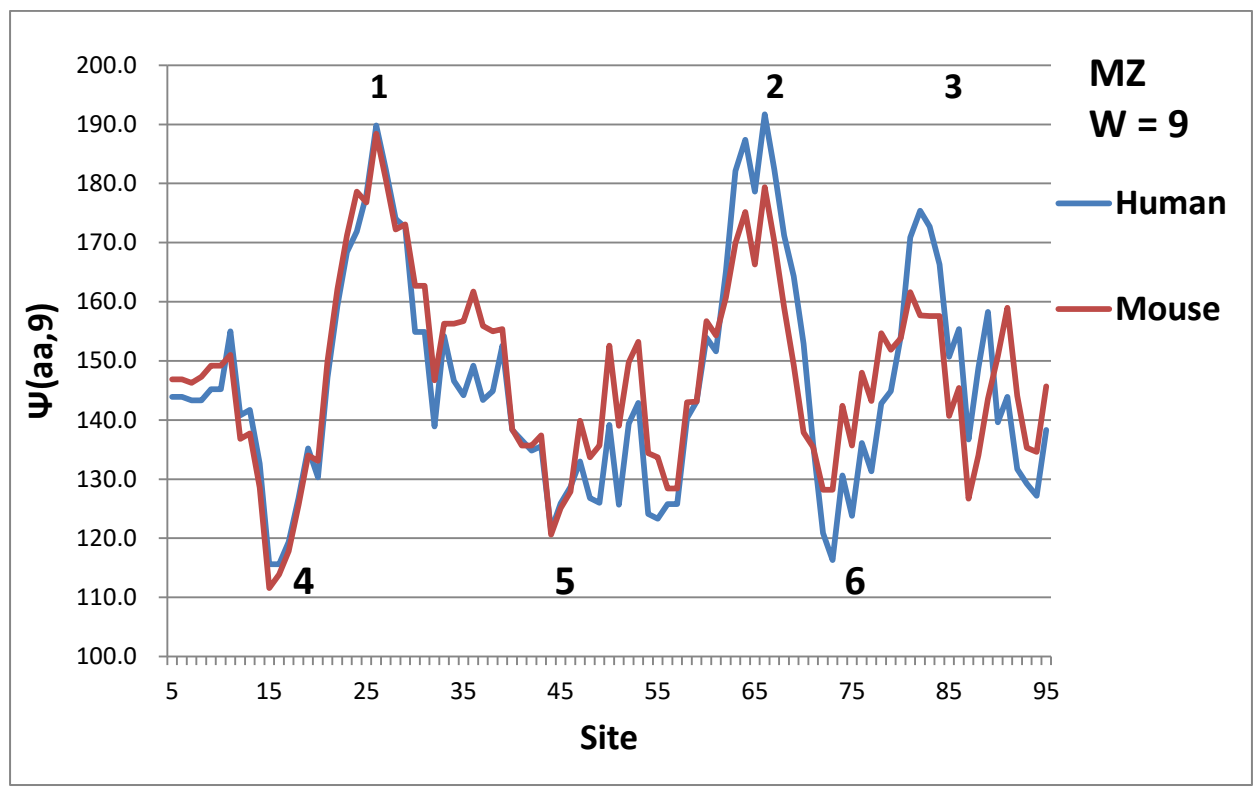

Fig. 2. Hydroprofiles of human and mouse, using the MZ (second-order phase transition) scale
[13] with W = 9. There are three hydrophobic edges (1 - 3), and 3 hydrophilic edges (4 - 6),

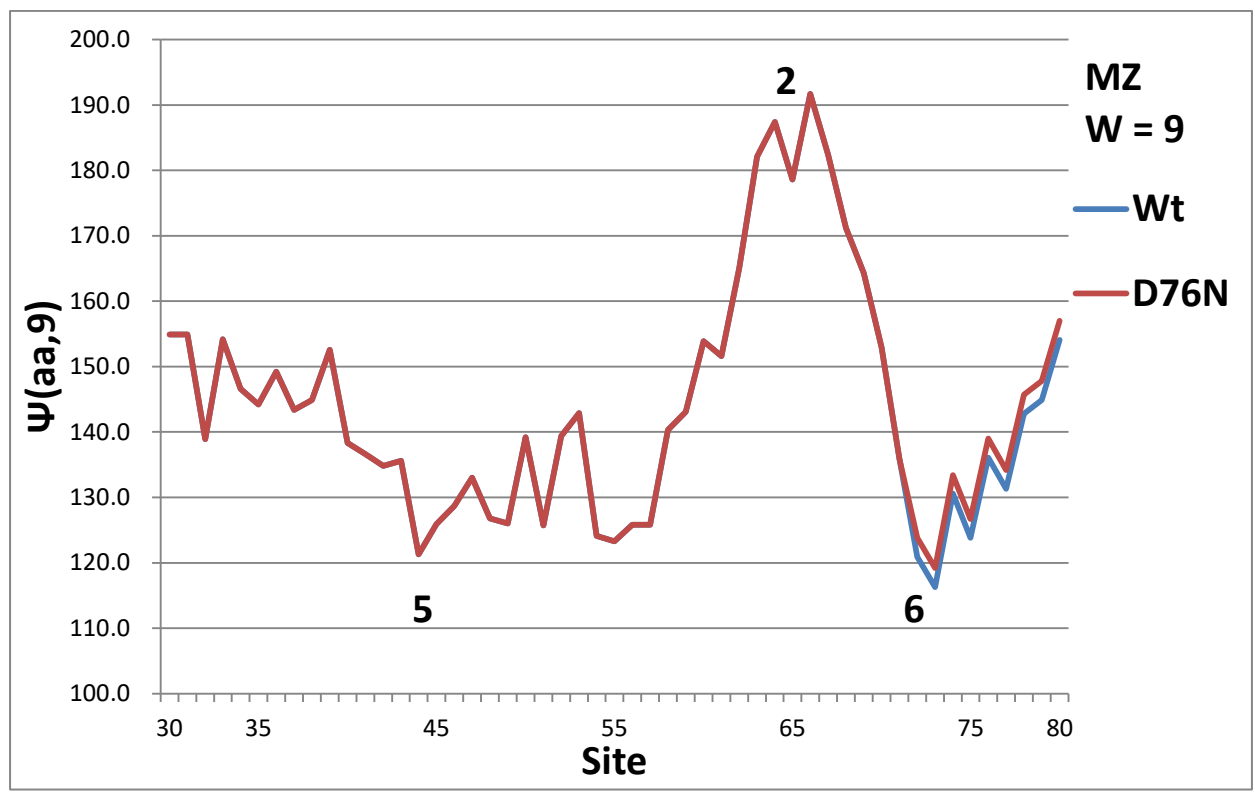

Fig. 3. The regions of Ψ(aa,9) including the hydrophilic edges 5 and 6 are enlarged in order to show the differences between the wild type human profile, and that with the mutation D76N. The mutation brings edge 6 into better alignment with edge 5.

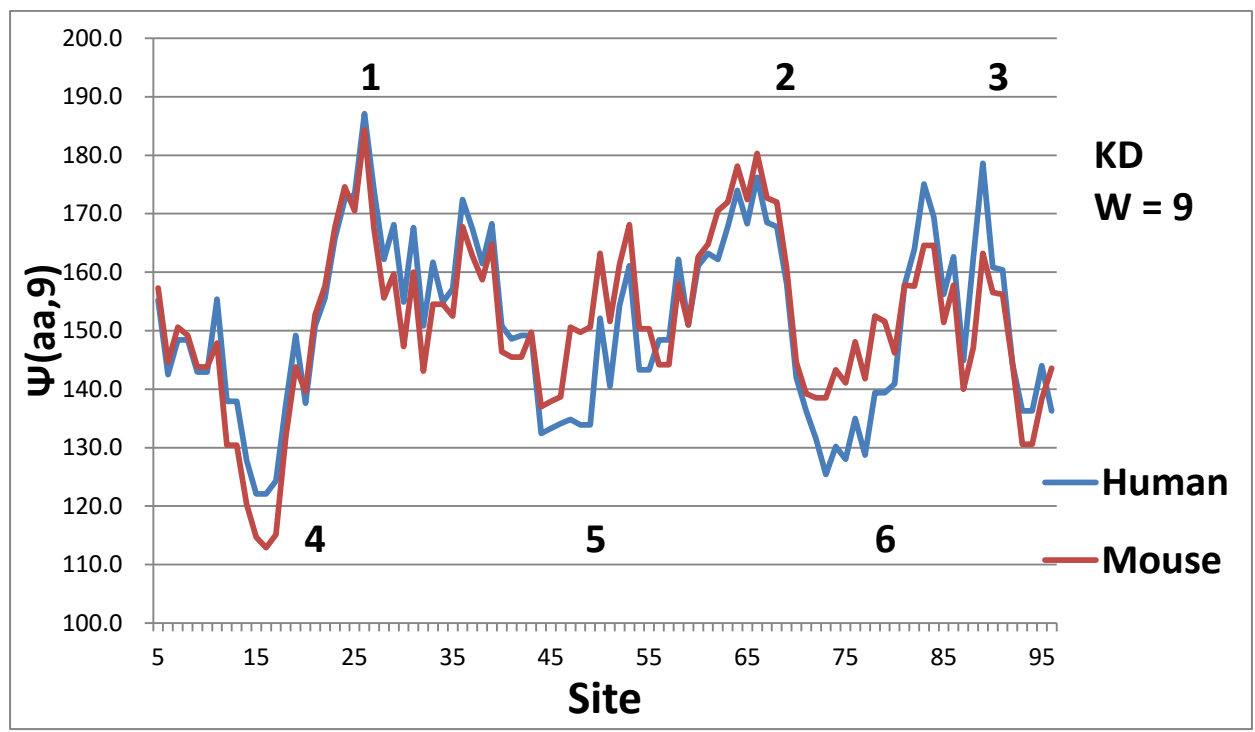

Fig. 4. This is similar to Fig. 2, but here the first-order KD scale [8,19] is used to compute Ψ(aa,9). The profiles are qualitatively similar to those with the second-order MZ scale in Fig. 2, but the edges (for example, 5 and 6) are not well aligned here.